\documentclass[prd,
eqsecnum,floatfix,a4paper,showpacs,superscriptaddress,groupedaddress,nofootinbib12pt]{revtex4}
\usepackage[utf8]{inputenc}

\usepackage{amsmath}
\usepackage{latexsym}
\usepackage{amssymb}
\usepackage{amsfonts}
\usepackage{mathtools}
\usepackage{bm}
\usepackage{amsthm}
\usepackage{graphicx}
\usepackage{color}
\usepackage{changes}
\usepackage[normalem]{ulem}
\usepackage{natbib}
\usepackage{appendix}
\usepackage{subcaption}
\usepackage{hyperref}
\begin{document}
\title{Possibility of Over-spinning Kerr Blackhole}
\author{Dishari Malakar\footnote{Current adddress: Missouri University of Science and Technology, Missouri S\&T, Rolla, MO 65409, U.S.A.}}
\email{dmkwf@umsystem.edu}
\author{K Rajesh  Nayak}
\email{rajesh@iiserkol.ac.in}
\affiliation{
The Center of Excellence in Space Sciences India (CESSI) and Department of Physical Sciences, Indian Institute of Science Education and Research Kolkata, Mohanpur,  West Bengal, 741246, India.}
\begin{abstract}
   We have developed a methodology to test the age-old cosmic censorship hypothesis in Kerr geometry. We have shown that the Kerr black hole can be overspun by particles captured from the innermost stable circular orbit. However, it appears that this does not happen for particles coming from infinity. We have also observed that overspinning becomes possible only when the black hole approaches a near-extremal limit, which approaches as we increase the black hole's mass. Our study demonstrates that an extremal black hole cannot be overspun. However, our methodology neglects backreaction and self-force effects, which could affect the overspinning.
\end{abstract}


\maketitle

\section{Introduction}
According to the Einstein's theory of relativity, matter undergoes a catastrophic convergence to a point-like region where both density of the matter and the curvature of the spacetime diverge to infinity - referred to as the singularity. Because of Raychaudhuri's equations within the framework of Einstein's general relativity, the singularities are often unavoidable~\cite{raychaudhuri1955relativistic}. The Cosmic Censorship Conjecture says that the singularity is hidden from outside region by an event horizon~\cite{penrose1969gravitational}. According to Penrose, there should exist some physical principle, which is not understood yet, that excludes naked singularities as the solutions to the equations of general relativity. When it is violated, the naked singularities are  exposed to afar. In the absence of a rigorous proof, we look for counter examples and for possible violation of the conjecture. One of such methods is by perturbing the blackhole with a test particle and look for over-spinning or over-charging of the same. In this paper, we have considered the Kerr spacetime around a rotating, uncharged, axially-symmetric blackhole. We say that the blackhole has over-spun when $a>M$ rendering the event horizon to be non-existent. Here, $M$ and $a$ are the blackhole's mass and angular momentum parameters. Similarly, we can describe the over-charging of charged blackholes like Reissner Nordstr\"{o}m.

The first approach of this kind was developed by Wald where he showed that it was not possible to  over-spin or over-charge an extremal blackhole when perturbed by test particles or fields~\cite{wald1974gedanken, wald1999gravitational}. Similar results were obtained by studying scalar and electromagnetic fields in extremal Kerr-Newman blackholes~\cite{semiz2011dyonic, toth2012test}. A recent study has also shown that it is not possible to destroy an extremal blackhole with test fields~\cite{natario2016test}. In recent times, Hubeny introduced the concept of a nearly extremal blackhole in the place of an extremal one and shown that it is possible to overcharge a near-extremal Reissner Nordstr\"{o}m blackhole~\cite{hubeny1999overcharging}. However, de Felice and Yu proved that an extremal Reissner Nordstr\"{o}m blackhole can be evolved to a naked singularity by  absorption of a neutral particle~\cite{de2001turning}. A quantum manifestation of the same was shown by Matsas and Silva~\cite{matsas2007overspinning}. Following Hubeny’s methodology, Jacobson and Sotiriou have argued that when a point test particle is brought into the Kerr blackhole, it can overspin the black hole with specific energy and angular momentum neglecting the backreaction and the self-force effects~\cite{jacobson2009overspinning}. The cosmic censorship gets restored when the backreaction effects were taken into account~\cite{hod2008weak, shaymatov2015destroying}. There have been many studies where Barausse et. al., Colleoni et. al., and Zimmerman et. al. have argued self force might act as the cosmic censorship\cite{barausse2010test, colleoni2015self, zimmerman2013self}. There are several other studies that have discussed the cosmic censorship in various spacetimes using classical as well as quantum gravity~\cite{ford1990moving, hod1999black, matsas2007overspinning, singh1999gravitational}. But, it has not yet been possible to attain a satisfying answer to this interesting problem.

In this paper, we have considered the Kerr spacetime to validate the cosmic censorship. We work closely along the line of  Jacobson  approach by taking Kerr blackholes with all possible rotational parameters~\cite{jacobson2009overspinning}. By using the test particle approximation, we accrete  a beam of particles into the blackhole at a steady rate. It is also assumed that the mass and angular momentum of the particles are absorbed by the blackhole after it is captured. In section~\ref{sec:over_spin}, we present a detailed formalism to  of our approach. In section~\ref{sec:infty} and~\ref{sec:isco}, two cases are considered - in the first case, we consider particles arriving from infinity are captured and  in the second case, blackhole captures particles from the innermost stable circular orbit (ISCO). These particles are absorbed by the blackhole which leads to a change in its mass and angular momentum. A critical factor is the ration of absorbed energy  and angular momentum, which determines the possibility of over-spinning.  Finally we close the article with  a brief concluding remarks in section~\ref{sec:conclusion}.
\section{Our approach to over-spin \label{sec:over_spin}}
Ted and Jacobson had shown in their work that it was possible to overspin a near extremal Kerr blackhole with test particle approximation~\cite{jacobson2009overspinning}. We extend their result by considering a wider class of Kerr blackholes that is, their mass and angular momentum ratio can range anywhere between 0 and 1. Our idea for overspinning starts with a blackhole of mass $M_0$ and angular momentum parameter $a_0$ inside which we send a ray of particles of energy $E$, rest mass $\mu$ and angular momentum $L$. We have assumed that the particles follow equatorial geodesics and cross the outer event horizon to fall inside the blackhole. Moreover, the particle is considered to be a test particle. We have also considered that the mass and the angular momentum of the particle are absorbed by the blackhole which in turn change the blackhole's parameters adiabatically.

The goal is to see if it is possible to make the angular momentum of the composite object greater than its mass. Thus, after the addition of $n$ particles, the solution should represent a Kerr spacetime with $M_n<a_n$, which implies the event horizon will cease to exist. Hence, we are left with a naked singularity. In this way, we can violate the weak cosmic censorship conjecture.
Our approach is based on pure classical general relativity ignoring the back reaction effects - we have not considered any quantum processes as done in some recent works~\cite{matsas2007overspinning}.
From here on, we will use the natural units i.e. $G=c=1$.
\subsection{Spin up Mechanism}
The Kerr metric in standard Boyer-Lindquist coordinates $(t,r,\theta, \phi)$ in $(-,+,+,+)$ notation,  is given by:
\begin{equation}
	ds^2 = -dt^2 +\Sigma \Big(\dfrac{dr^2}{\Delta}+d\theta^2\Big) +\Big(r^2+a^2 \Big)sin^2\theta d\phi^2 +\dfrac{2Mr}{\Sigma}\Big(asin^2\theta d\phi-dt\Big)^2 \ , \label{k_m}
\end{equation}
where, $\Delta=r^2+a^2-2Mr$ and $\Sigma=r^2+a^2cos^2\theta$. The metric describes a stationary axially-symmetric vacuum solution, with mass parameter, $M$ and angular momentum per unit mass, $a$.  
The Kerr spacetime allows a timelike Killing vector, $\xi^a=\left(1, \, 0,\, 0, \, 0\right) $ and  spacelike Killing vector $\eta^a=\left( 0, \, 0, \, 0, \, 1\right)$.  The timelike Killing vector $\xi^a$ is not surface forming and hence define  a vector field,  $\chi_a$, which is timelike hyper-surface 
orthogonal. 
\begin{equation}
\chi_a = \xi_a - \frac{\xi^b\eta_b}{\eta^c\eta_c} \eta_a\, 
\end{equation}

 The  $\chi_a$ becomes Killing on the null surface and hence  it is called the horizon generating Killing vector.
 The horizon is given by the condition, $\Delta=0$, and explicit solution for the radial position in terms of $M$ and $a$ are  given by, $r_\pm = M \pm \sqrt{M^2-a^2}$. Clearly, the inner and the outer horizons,  $r_{+}$ and $r_{-}$  merge as $M \rightarrow a$ and  the event horizon seize to exist for $a>M$ with  the naked singularity exposed to outside observer.

When the particle goes inside the horizon the  mass and the angular momentum it carries with it in to the blackhole  is related to the stress energy tensor, $T^{ab}$,  of the particle via:
\begin{eqnarray}
    \delta M =  \lim_{r \rightarrow r_{+}} \, \int T^{ab}\chi_a d\Sigma_b \ , \quad
     \delta J =\lim_{r \rightarrow r_{+}} \, \Omega_H \int T^{ab}\eta_a d\Sigma_b \label{i2} \ .
\end{eqnarray}
In the above equation, $\Omega_H=-a/2Mr_+$ is the angular velocity of the event horizon. And the surface element of null hypersurface:
\begin{equation}
	d\Sigma_a= \chi_a \sqrt{ \dfrac{(r^2+a^2)^2-\Delta a^2 sin^2\theta}{\Sigma} } d\phi. \label{g0}
\end{equation}
We assume that the in-going particle is moving along equatorial plane. 
 These expressions provide the flux for a single particle. By the method of induction, we compute the total mass and angular momentum of the blackhole due to $n$ number of particles coming from infinity and ISCO as well as the possibility to overspin the blackhole in the following sections.
\subsection{Stress Energy Tensor:} We start with stress energy tensor for particles moving along a geodesic. The stress energy tensor geodesic is approximated by  perfect fluid with no pressure. In terms of the four-velocity, $u^a$ and rest mass $\mu$, the we have, $T^{ab}=\mu \, u^a u^b$.  The  four-velocity, $u^a \, = \left(\dot t, \,  \dot r, \, 0, \, \dot \phi \right)$, for  geodesic  motion in Kerr spacetime is described in terms of conserved  quantities, energy $E$ and angular momentum $L$:
\begin{eqnarray}
	\dot{t} &= &\dfrac{1}{\mu \Delta}[(r^2+a^2+\dfrac{2Ma^2}{r})E-\dfrac{2Ma}{r}L] \ , \nonumber \\
	\dot{\phi} &=& \dfrac{1}{\mu \Delta}[(1-\dfrac{2M}{r})L+\dfrac{2Ma}{r}E] \ , \nonumber \\
	\mu^2 r^2\dot{r}^2 &= &-\mu^2\Delta+r^2 E^2 +\dfrac{2M}{r}(L-aE)^2-(L^2-a^2E^2). \label{eq:geodesic}
\end{eqnarray}
Note that, for simplicity, here we take $\theta=\frac{\pi}{2}$, $\dot\theta=0$ and $\ddot \theta=0$.
\section{Particles from Infinity \label{sec:infty}}
In this section, we consider particles with rest mass $\mu$, energy $E$ and angular momentum $L$ following equatorial geodesics fall into the blackhole. It is important to note that not all geodesics coming from infinity enter into the blackhole. We consider the maximum angular momentum of the particle ($L=aE)$ for which it crosses the event horizon~\cite{chandrasekhar1998mathematical}. Using this condition, we find the non-zero components of the stress-energy tensor from the equations of motion(\ref{eq:geodesic}). Thus, the mass and angular momentum flux carried into the blackhole by one particle per unit angle is computed using equations (\ref{i2}):
\begin{eqnarray}
	\delta M  =  \dfrac{\pi E^2r_+^2}{\mu M} \ , \quad
	\delta J= \dfrac{\pi J^2E^2}{\mu M^3} \ .
\end{eqnarray}
By the method of induction, the final mass and angular momentum of the blackhole is found out after $n$ particles enter into the blackhole. 
\begin{eqnarray}
	M_n &=& M_{n-1} + \dfrac{E^2r_{(n-1)+}^2}{2\mu M_{n-1}} \ ,  \nonumber \\ 
	a_n &=& a_{n-1} + \dfrac{E^2 a_{n-1} }{2\mu M_{n-1}^2}\Big(a_{n-1}-r_{(n-1)+}^2\Big) \ . \label{a7}
\end{eqnarray}

\begin{figure}[t]
	\centering
	\includegraphics[scale=0.3]{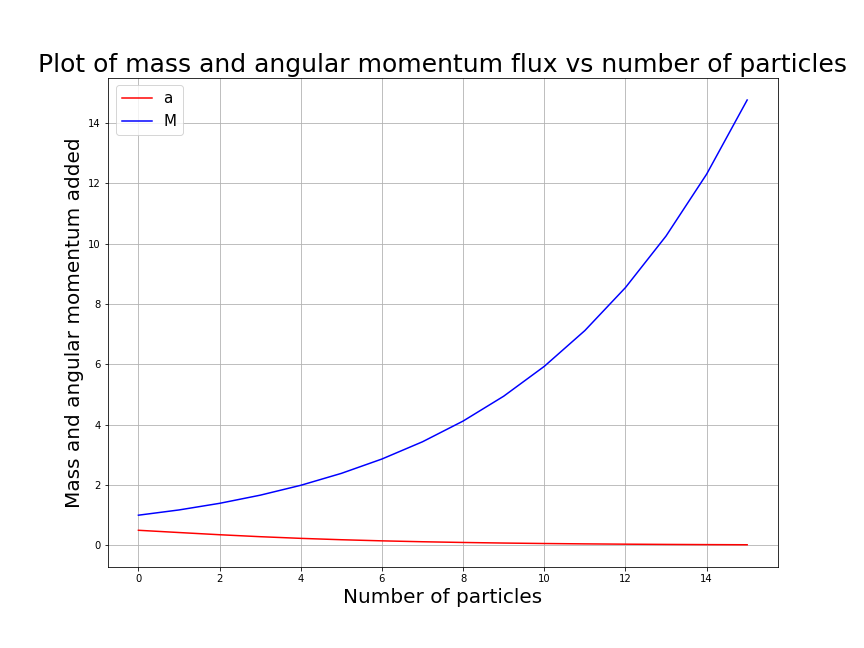}
	\caption{Plot of Mass (M) and angular momentum (a) added to Kerr blackhole for $n$ particles coming from infinity. $M_0=1.0$, $a_0=0.5$ and $E=\mu=0.01$ }
	\label{fig}
\end{figure}

Note that, here we calculate $a_n=J_n/M_n$ by binomial expansion. These equations are plotted against the number of particles and it is seen that the angular momentum parameter $a$ decreases with the entry of more and more particles, while mass of the blackhole increases as shown in the figure-\ref{fig}. Thus, there is no such region where $a$ overtakes $M$, making overspinning impossible with particles coming from infinity. 

\section{Particles from ISCO \label{sec:isco}}
In this section, we investigate the possibility of blackhole over-spinning   by absorbing particles moving along inner-most stable orbit. Once again we confine to the  equatorial orbits. For such orbits, the effective radial potential is given by : \cite{potential}
\begin{equation}
	V(r)=(1-E^2)r^4-2Mr^3+[a^2(1-E^2)+L_z^2]r^2-2M(aE-L_z)^2 r.
\end{equation}
Here, $E=E/\mu$ ie. energy per unit mass and $L_z=L_z/\mu$, angular momentum per unit mass of the particle.  The  $E$ and $L_z$ are given by:
\begin{eqnarray}
	E &= &\dfrac{1-\frac{2M}{r}+\frac{aM^{1/2}}{r^{3/2}}}{\left[1-\frac{3M}{r}+\frac{2aM^{1/2}}{r^{3/2}}\right]^{1/2}} \, ,  \nonumber \\
	L_z &= & \dfrac{M^{1/2}r^{1/2}-\frac{2aM}r+\frac{a^2M^{1/2}}{r^{3/2}}}{\left[1-\frac{3M}r+\frac{2aM^{1/2}}{r^{3/2}}\right]^{1/2}} \label{EL}\ .
\end{eqnarray}
We obtain the location of the innermost stable circular orbit by solving $dE/dr=0$, we get:
\begin{equation}
	1-\dfrac{6M}{r}+\dfrac{8aM^{1/2}}{r^{3/2}}-\dfrac{3a^2}{r^2}=0 \ . \label{min_e}
\end{equation}
The value of $r$ satisfying equation (\ref{min_e}) gives the minimum energy stable orbit around the black hole. This is also the minimum angular momentum stable orbit.  We  use equation (\ref{min_e}) to find the ISCO which in turn will provide the energy and angular momentum being added to the blackhole from equation (\ref{EL}).

 From equation~(\ref{EL}), It can be seen clearly that the energy of the particle reaches a saturation at 1,  the angular momentum of the particle at ISCO increases for increasing mass.
With the above equations, we  compute the mass and angular momentum flux into the blackhole for a single  particle:
\begin{eqnarray}
	\delta M=\dfrac{1}{2M}\left(2ME-\dfrac{aL}{r_+}\right)^2 \ , \quad \delta J= \dfrac{JL}{2M^2r_+}\left(2ME-\dfrac{JL}{Mr_+}\right) \ .
\end{eqnarray}
Hence the recursion relation for $M_n$ and $a_n=J_n/M_n$ after $n$ particles are added, is given by:
\begin{eqnarray}
	M_n &=& M_{n-1}+\dfrac{1}{2M_{n-1}}\left[2M_{n-1}E_{n-1}-\dfrac{a_{n-1}L_{n-1}}{r_{+(n-1)}}\right]^2 \ , \nonumber \\
	a_n &=& a_{n-1} +\dfrac{a_{n-1}}{2M_{n-1}^2}\left[2M_{n-1}E_{n-1}-\dfrac{a_{n-1}L_{n-1}}{r_{+(n-1)}}\right] \left[\dfrac{L_{n-1}}{r_{+(n-1)}}\left(1+a_{n-1}\right)-2M_{n-1}E_{n-1}\right] \ . \label{a_n}
\end{eqnarray}
We numerically find out  the optimum value of initial angular momentum parameter, $a_{i0}$,  depending on the mass of the blackhole, for which it over-spins. Below this optimum value, the angular momentum of the blackhole decreases with the entry of the particles, whereas the mass goes on increasing. Thus, over-spinning could not be achieved for such blackholes. We also see that for the extremal case i.e. $M_i=a_i$, the blackhole doesn't over-spin as the ISCO coincides with the horizon. 

For $M_i=1.0$, the blackhole over-spins for $a_{i0}=0.76632$ with the entry of the 20th particle. For all the values of $a_i$ ranging between $a_i=a_{i0}$ and $a_i\rightarrow M_i$, {\it i.e}, nearly extremal case, the blackhole does over-spin. It has also been seen that as the initial mass of the blackhole increases, the optimum value of initial angular momentum increases and $a_{i0}/M_i \rightarrow 1$. Thus,  with increasing mass, only the near extremal blackholes can be overspun.

\begin{figure}
	\centering
	\begin{subfigure}{0.3\textwidth}
		\includegraphics[width=\textwidth]{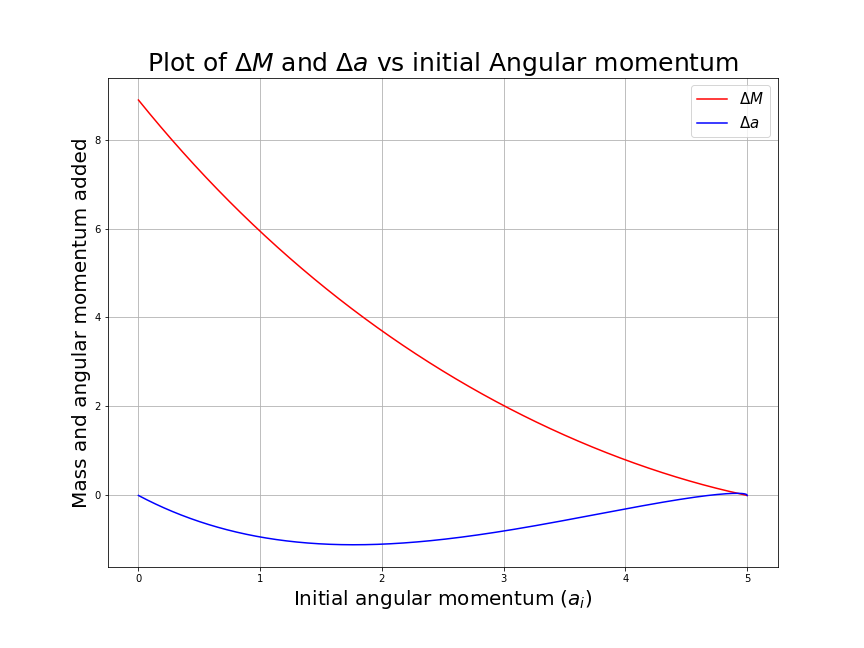}
		\caption{For $M_i=5.0$}
		\label{fig4b}
	\end{subfigure}
	\begin{subfigure}{0.3\textwidth}
		\includegraphics[width=\textwidth]{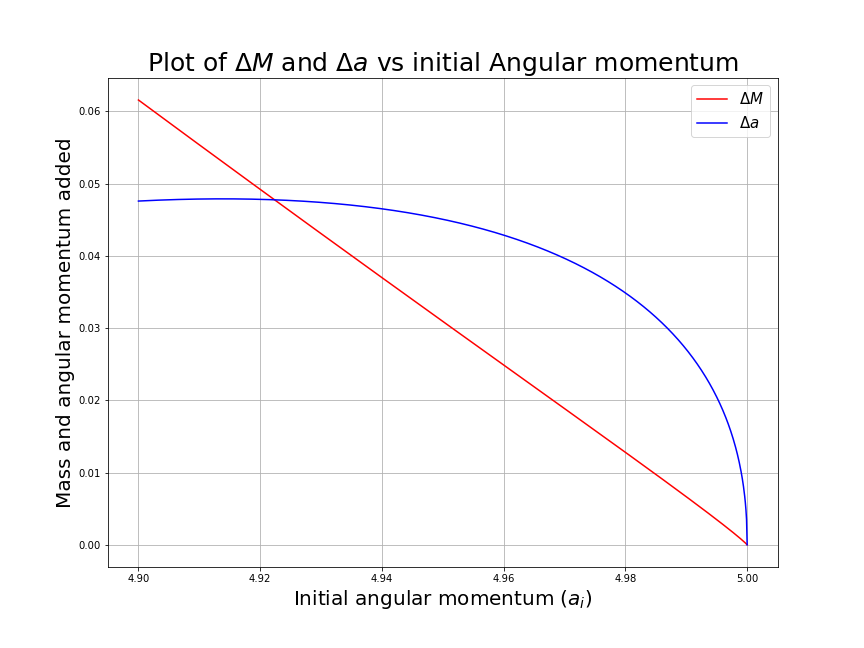}
		\caption{$a_i=4.90$ to $a_i=5.0$}
		\label{fig4c}
	\end{subfigure}
	\begin{subfigure}{0.3\textwidth}
		\includegraphics[width=\textwidth]{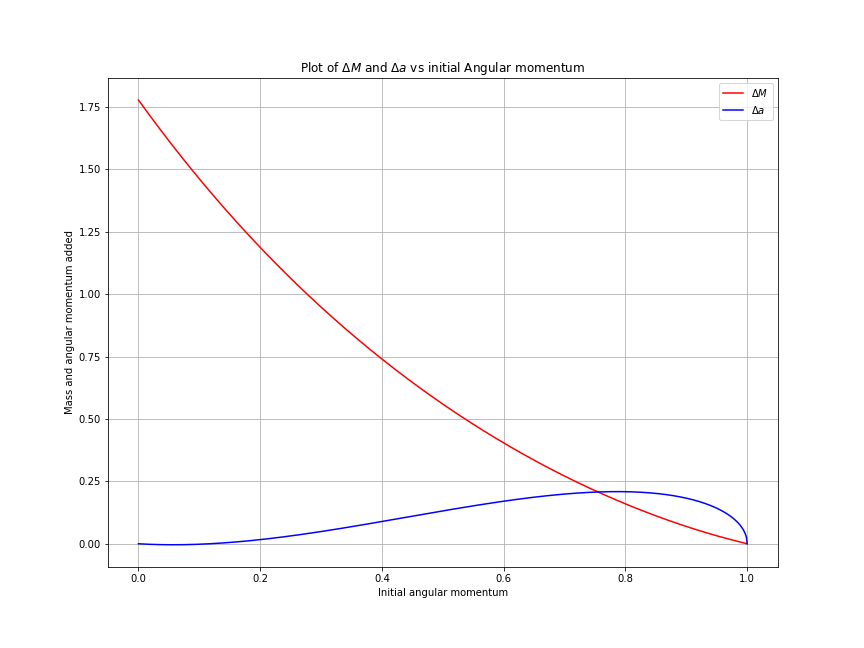}
		\caption{For $M_i=1.0$}
		\label{fig4a}
	\end{subfigure}
	
	\caption{Plot of $\Delta M$ and $\Delta a$ with initial angular momentum for different $M_i$}
	\label{fig4}
\end{figure}

To see the reason behind this, we have plotted the change in angular momentum ($\Delta a$) and the change in mass ($\Delta M$) against the initial angular momentum. The result is interesting. At low $a_i$, the value of $\Delta a$ is less than that of $\Delta M$ and it increases at a point near the optimum angular momentum. Also we see that as mass of the blackhole is increased, the crossing point of $\Delta a$ and $\Delta M$ reaches to the near extremal value [Figure \ref{fig4}]. Thus, with our methodology, we have been able to overspin the Kerr blackhole when it captures the particles from ISCO.

\section{Conclusion \label{sec:conclusion}}

In this work, we have tried to violate the weak cosmic censorship conjecture by throwing in particles into the Kerr Blackhole. We have considered two cases, where in one part, the particles are coming from infinity and in second part, the particles are captured by the blackhole from the innermost stable circular orbit. It was observed that the particles from infinity could not overspin the Kerr blackhole, whereas the particles from ISCO were successful in doing so. 

The results from ISCO part are pretty interesting. As the initial mass of the blackhole increases, the chances of overspinning it go to the near extremal limit, as suggested by Ted and Jacobson~\cite{jacobson2009overspinning}. Moreover, the extremal blackhole could not be overspun, agreeing to Wald's hypothesis~\cite{wald1974gedanken}. 

In our work, we have considered the test particle approximation which nullifies the back reaction effects to some degree. But previous studies have already shown that self force effects can indeed impose the cosmic censorship~\cite{barausse2010test, colleoni2015self}. Moreover, we do not know how the particle might behave inside the event horizon - our assumption that the captured particle's mass and angular momentum are directly absorbed by the blackhole might not be true in all cases. A detailed study keeping these in mind can be done to test the conjecture more thoroughly.

\section{Acknowledgment}

DM would like to thank IISER Kolkata and CESSI for the hospitality during the work.

\bibliography{reference}

\begin{thebibliography}{22}
\expandafter\ifx\csname natexlab\endcsname\relax\def\natexlab#1{#1}\fi
\expandafter\ifx\csname bibnamefont\endcsname\relax
  \def\bibnamefont#1{#1}\fi
\expandafter\ifx\csname bibfnamefont\endcsname\relax
  \def\bibfnamefont#1{#1}\fi
\expandafter\ifx\csname citenamefont\endcsname\relax
  \def\citenamefont#1{#1}\fi
\expandafter\ifx\csname url\endcsname\relax
  \def\url#1{\texttt{#1}}\fi
\expandafter\ifx\csname urlprefix\endcsname\relax\def\urlprefix{URL }\fi
\providecommand{\bibinfo}[2]{#2}
\providecommand{\eprint}[2][]{\url{#2}}

\bibitem[{\citenamefont{Raychaudhuri}(1955)}]{raychaudhuri1955relativistic}
\bibinfo{author}{\bibfnamefont{A.}~\bibnamefont{Raychaudhuri}},
  \bibinfo{journal}{Phys. Rev.} \textbf{\bibinfo{volume}{98}},
  \bibinfo{pages}{1123} (\bibinfo{year}{1955}).

\bibitem[{\citenamefont{Penrose}(1969)}]{penrose1969gravitational}
\bibinfo{author}{\bibfnamefont{R.}~\bibnamefont{Penrose}},
  \bibinfo{journal}{Nuovo Cimento} \textbf{\bibinfo{volume}{1}},
  \bibinfo{pages}{252} (\bibinfo{year}{1969}).

\bibitem[{\citenamefont{Wald}(1974)}]{wald1974gedanken}
\bibinfo{author}{\bibfnamefont{R.}~\bibnamefont{Wald}}, \bibinfo{journal}{Ann.
  Phys. (N.Y.)} \textbf{\bibinfo{volume}{82}}, \bibinfo{pages}{548}
  (\bibinfo{year}{1974}).

\bibitem[{\citenamefont{Wald}(1999)}]{wald1999gravitational}
\bibinfo{author}{\bibfnamefont{R.~M.} \bibnamefont{Wald}}, in
  \emph{\bibinfo{booktitle}{Black holes, gravitational radiation and the
  universe}} (\bibinfo{publisher}{Springer}, \bibinfo{year}{1999}), pp.
  \bibinfo{pages}{69--86}.

\bibitem[{\citenamefont{Semiz}(2011)}]{semiz2011dyonic}
\bibinfo{author}{\bibfnamefont{I.}~\bibnamefont{Semiz}}, \bibinfo{journal}{Gen.
  Relativ. Gravit.} \textbf{\bibinfo{volume}{43}}, \bibinfo{pages}{833}
  (\bibinfo{year}{2011}).

\bibitem[{\citenamefont{T{\'o}th}(2012)}]{toth2012test}
\bibinfo{author}{\bibfnamefont{G.~Z.} \bibnamefont{T{\'o}th}},
  \bibinfo{journal}{Gen. Relativ. Gravit.} \textbf{\bibinfo{volume}{44}},
  \bibinfo{pages}{2019} (\bibinfo{year}{2012}).

\bibitem[{\citenamefont{Nat{\'a}rio et~al.}(2016)\citenamefont{Nat{\'a}rio,
  Queimada, and Vicente}}]{natario2016test}
\bibinfo{author}{\bibfnamefont{J.}~\bibnamefont{Nat{\'a}rio}},
  \bibinfo{author}{\bibfnamefont{L.}~\bibnamefont{Queimada}}, \bibnamefont{and}
  \bibinfo{author}{\bibfnamefont{R.}~\bibnamefont{Vicente}},
  \bibinfo{journal}{Class. Quantum Grav.} \textbf{\bibinfo{volume}{33}},
  \bibinfo{pages}{175002} (\bibinfo{year}{2016}).

\bibitem[{\citenamefont{Hubeny}(1999)}]{hubeny1999overcharging}
\bibinfo{author}{\bibfnamefont{V.~E.} \bibnamefont{Hubeny}},
  \bibinfo{journal}{Phys. Rev. D} \textbf{\bibinfo{volume}{59}},
  \bibinfo{pages}{064013} (\bibinfo{year}{1999}).

\bibitem[{\citenamefont{de~Felice and Yunqiang}(2001)}]{de2001turning}
\bibinfo{author}{\bibfnamefont{F.}~\bibnamefont{de~Felice}} \bibnamefont{and}
  \bibinfo{author}{\bibfnamefont{Y.}~\bibnamefont{Yunqiang}},
  \bibinfo{journal}{Class. Quantum Grav.} \textbf{\bibinfo{volume}{18}},
  \bibinfo{pages}{1235} (\bibinfo{year}{2001}).

\bibitem[{\citenamefont{Matsas and Da~Silva}(2007)}]{matsas2007overspinning}
\bibinfo{author}{\bibfnamefont{G.~E.} \bibnamefont{Matsas}} \bibnamefont{and}
  \bibinfo{author}{\bibfnamefont{A.~R.} \bibnamefont{Da~Silva}},
  \bibinfo{journal}{Phys. Rev. Lett.} \textbf{\bibinfo{volume}{99}},
  \bibinfo{pages}{181301} (\bibinfo{year}{2007}).

\bibitem[{\citenamefont{Jacobson and
  Sotiriou}(2009)}]{jacobson2009overspinning}
\bibinfo{author}{\bibfnamefont{T.}~\bibnamefont{Jacobson}} \bibnamefont{and}
  \bibinfo{author}{\bibfnamefont{T.~P.} \bibnamefont{Sotiriou}},
  \bibinfo{journal}{Phys. Rev. Lett.} \textbf{\bibinfo{volume}{103}},
  \bibinfo{pages}{141101} (\bibinfo{year}{2009}).

\bibitem[{\citenamefont{Hod}(2008)}]{hod2008weak}
\bibinfo{author}{\bibfnamefont{S.}~\bibnamefont{Hod}}, \bibinfo{journal}{Phys.
  Rev. Lett.} \textbf{\bibinfo{volume}{100}}, \bibinfo{pages}{121101}
  (\bibinfo{year}{2008}).

\bibitem[{\citenamefont{Shaymatov et~al.}(2015)\citenamefont{Shaymatov, Patil,
  Ahmedov, and Joshi}}]{shaymatov2015destroying}
\bibinfo{author}{\bibfnamefont{S.}~\bibnamefont{Shaymatov}},
  \bibinfo{author}{\bibfnamefont{M.}~\bibnamefont{Patil}},
  \bibinfo{author}{\bibfnamefont{B.}~\bibnamefont{Ahmedov}}, \bibnamefont{and}
  \bibinfo{author}{\bibfnamefont{P.~S.} \bibnamefont{Joshi}},
  \bibinfo{journal}{Phys. Rev. D} \textbf{\bibinfo{volume}{91}},
  \bibinfo{pages}{064025} (\bibinfo{year}{2015}).

\bibitem[{\citenamefont{Barausse et~al.}(2010)\citenamefont{Barausse, Cardoso,
  and Khanna}}]{barausse2010test}
\bibinfo{author}{\bibfnamefont{E.}~\bibnamefont{Barausse}},
  \bibinfo{author}{\bibfnamefont{V.}~\bibnamefont{Cardoso}}, \bibnamefont{and}
  \bibinfo{author}{\bibfnamefont{G.}~\bibnamefont{Khanna}},
  \bibinfo{journal}{Phys. Rev. Lett.} \textbf{\bibinfo{volume}{105}},
  \bibinfo{pages}{261102} (\bibinfo{year}{2010}).

\bibitem[{\citenamefont{Colleoni et~al.}(2015)\citenamefont{Colleoni, Barack,
  Shah, and Van De~Meent}}]{colleoni2015self}
\bibinfo{author}{\bibfnamefont{M.}~\bibnamefont{Colleoni}},
  \bibinfo{author}{\bibfnamefont{L.}~\bibnamefont{Barack}},
  \bibinfo{author}{\bibfnamefont{A.~G.} \bibnamefont{Shah}}, \bibnamefont{and}
  \bibinfo{author}{\bibfnamefont{M.}~\bibnamefont{Van De~Meent}},
  \bibinfo{journal}{Phys. Rev. D} \textbf{\bibinfo{volume}{92}},
  \bibinfo{pages}{084044} (\bibinfo{year}{2015}).

\bibitem[{\citenamefont{Zimmerman et~al.}(2013)\citenamefont{Zimmerman, Vega,
  Poisson, and Haas}}]{zimmerman2013self}
\bibinfo{author}{\bibfnamefont{P.}~\bibnamefont{Zimmerman}},
  \bibinfo{author}{\bibfnamefont{I.}~\bibnamefont{Vega}},
  \bibinfo{author}{\bibfnamefont{E.}~\bibnamefont{Poisson}}, \bibnamefont{and}
  \bibinfo{author}{\bibfnamefont{R.}~\bibnamefont{Haas}},
  \bibinfo{journal}{Phys. Rev. D} \textbf{\bibinfo{volume}{87}},
  \bibinfo{pages}{041501} (\bibinfo{year}{2013}).

\bibitem[{\citenamefont{Ford and Roman}(1990)}]{ford1990moving}
\bibinfo{author}{\bibfnamefont{L.}~\bibnamefont{Ford}} \bibnamefont{and}
  \bibinfo{author}{\bibfnamefont{T.~A.} \bibnamefont{Roman}},
  \bibinfo{journal}{Phys. Rev. D} \textbf{\bibinfo{volume}{41}},
  \bibinfo{pages}{3662} (\bibinfo{year}{1990}).

\bibitem[{\citenamefont{Hod}(1999)}]{hod1999black}
\bibinfo{author}{\bibfnamefont{S.}~\bibnamefont{Hod}}, \bibinfo{journal}{Phys.
  Rev. D} \textbf{\bibinfo{volume}{60}}, \bibinfo{pages}{104031}
  (\bibinfo{year}{1999}).

\bibitem[{\citenamefont{Singh}(1999)}]{singh1999gravitational}
\bibinfo{author}{\bibfnamefont{T.~P.} \bibnamefont{Singh}}, \bibinfo{journal}{J
  Astrophys Astron} \textbf{\bibinfo{volume}{20}}, \bibinfo{pages}{221}
  (\bibinfo{year}{1999}).

\bibitem[{\citenamefont{Lehner et~al.}(2016)\citenamefont{Lehner, Myers,
  Poisson, and Sorkin}}]{lehner2016gravitational}
\bibinfo{author}{\bibfnamefont{L.}~\bibnamefont{Lehner}},
  \bibinfo{author}{\bibfnamefont{R.~C.} \bibnamefont{Myers}},
  \bibinfo{author}{\bibfnamefont{E.}~\bibnamefont{Poisson}}, \bibnamefont{and}
  \bibinfo{author}{\bibfnamefont{R.~D.} \bibnamefont{Sorkin}},
  \bibinfo{journal}{Phys. Rev. D} \textbf{\bibinfo{volume}{94}},
  \bibinfo{pages}{084046} (\bibinfo{year}{2016}).

\bibitem[{\citenamefont{Chandrasekhar}(1998)}]{chandrasekhar1998mathematical}
\bibinfo{author}{\bibfnamefont{S.}~\bibnamefont{Chandrasekhar}},
  \emph{\bibinfo{title}{The mathematical theory of black holes}},
  vol.~\bibinfo{volume}{69} (\bibinfo{publisher}{Oxford University Press},
  \bibinfo{year}{1998}).

\bibitem[{\citenamefont{Hirata}()}]{potential}
\bibinfo{author}{\bibfnamefont{C.~M.} \bibnamefont{Hirata}},
  \emph{\bibinfo{title}{Kerr black holes: Ii. precession, circular orbits, and
  stability}},
  \urlprefix\url{http://www.tapir.caltech.edu/~chirata/ph236/2011-12/lec27.pdf}.

\end{thebibliography}

\end{document}